\documentclass[useAMS,usenatbib]{mn2e}
\usepackage{epsfig}

\title[Binary Star Origin of High Field Magnetic White Dwarfs]
{Binary Star Origin of High Field Magnetic White Dwarfs}

\author[C. A. Tout, D. T. Wickramasinghe, J. Liebert,
L. Ferrario and J. E. Pringle]{C. A. Tout$^1$, D. T. Wickramasinghe$^2$, J. Liebert$^3$,
L. Ferrario$^2$ and J. E. Pringle$^1$\\
$^1$Institute of Astronomy, The Observatories, Madingley Road,
Cambridge CB3 0HA\\
$^2$The Australian National University, ACT 0200, Australia\\
$^3$Steward Observatory, University of Arizona, Tucson, AZ 85726, USA}

\begin{document}

\date{Accepted.  Received ; in original form} 

\pagerange{\pageref{firstpage}--\pageref{lastpage}} \pubyear{}

\maketitle

\label{firstpage}

\begin{abstract}
White dwarfs with surface magnetic fields in excess of $1\,$MG are
found as isolated single stars and relatively more often in magnetic
cataclysmic variables.  Some $1{,}253$ white dwarfs with a detached
low-mass main-sequence companion are identified in the Sloan Digital
Sky Survey but none of these is observed to show evidence for Zeeman
splitting of hydrogen lines associated with a magnetic field in excess
of $1\,$MG.  If such high magnetic fields on white dwarfs result from
the isolated evolution of a single star then there should be the same
fraction of high field white dwarfs among this SDSS binary sample as
among single stars.  Thus we deduce that the origin of such high
magnetic fields must be intimately tied to the formation of
cataclysmic variables.  The formation of a CV must involve orbital
shrinkage from giant star to main-sequence star dimensions.  It is
believed that this shrinkage occurs as the low-mass companion and the
white dwarf spiral together inside a common envelope.  CVs emerge as
very close but detached binary stars that are then brought together by
magnetic braking or gravitational radiation.  We propose that the
smaller the orbital separation at the end of the common envelope
phase, the stronger the magnetic field.  The magnetic cataclysmic
variables originate from those common envelope systems that almost
merge.  We propose further that those common envelope systems that
do merge are the progenitors of the single high field white dwarfs.
Thus all highly magnetic white dwarfs, be they single stars or the
components of MCVs, have a binary origin.  This hypothesis also
accounts for the relative dearth of single white dwarfs with fields of
$10^4 - 10^6\,$G.  Such intermediate-field white dwarfs are found
preferentially in cataclysmic variables.  In addition the bias towards
higher masses for highly magnetic white dwarfs is expected if a
fraction of these form when two degenerate cores merge in a common
envelope.  Similar scenarios may account for very high field neutron
stars.  From the space density of single highly magnetic white dwarfs
we estimate that about three times as many common envelope events lead
to a merged core as to a cataclysmic variable.
\end{abstract}


\section {Introduction}
White dwarfs appear to form with a wide range of surface magnetic
fields up to $10^9\,$G \citep{schmidt2003}.  Isolated white dwarfs can
be separated into two groups, those with high magnetic fields stronger
than $10^6\,$G (HFMWDs) and the rest with lower magnetic fields,
typically less than $10^5\,$G.  About ten per cent of isolated WDs are
HFMWDs \citep{liebert2005,kawka2007}.  A white dwarf with a close
companion that is overflowing its Roche lobe is a cataclysmic variable
\citep{warner1995}.  Among the cataclysmic variables about twenty-five
per cent \citep{wickramasinghe2000} have white dwarfs that are very
magnetic.  These fall into two classes.  Those with a highly magnetic
accreting primary star are called polars or AM~Herculis systems.  In
these systems the magnetosphere of the primary is able to totally
control the accretion flow from the secondary, such that no accretion
disc forms and the pair are locked in synchronous rotation at the
orbital period. The polars have measured fields in the range $10^7 -
10^8\,$G.  Of slightly weaker field are the intermediate polars
or DQ~Her systems in which the magnetic field of the primary does not
entirely prevent the formation of an accretion disc.  The white dwarf
is spun up to a rotation period shorter than the orbital period.  The
majority are deduced to have primaries with smaller
magnetic field strengths than the polars ($5 \times 10^6 - 10^7\,$G)
but the lower limit is uncertain.  There is some overlap at the high
field end with the polars.

In the Sloan Digital Sky Survey (SDSS) Data Release Five, some $1{,}253$
spectroscopically-observed close binary systems comprising a white
dwarf and a nondegenerate, usually M, dwarf (hereinafter WD+M) have
been identified and catalogued by \citet{silvestri2007}.  Of all these
pairs, none has been found in which the white dwarf is observed to
have a magnetic field above $2-3\,$MG, the limit of detectability via
the Zeeman splitting of the Balmer lines at the spectral resolutions
used.  Here we extend the size of the sample and strengthen greatly
the statistics on the apparent absence of magnetic white dwarf
pairings with a nondegenerate dwarf star, as first discussed at length
by \citet{liebert2005}.

Previously the best explanation for the apparent absence of magnetic
white dwarfs associated with M dwarfs put forth by \citet{liebert2005}
was related to the evidence that at least some strongly magnetic white
dwarfs have higher than average masses and smaller radii.  As
demonstrated by representative simulations of \citet{silvestri2007},
the smaller radius and luminosity of a magnetic primary causes its
spectrum to be masked more easily by an M companion.  However, recent
studies (see section~\ref{masses}) have shown that the HFMWDs are of a
range of masses with several of $0.6\,\rm M_{\odot}$ or less.  Thus,
although the mean mass for the HFMWDs is higher, it is not
sufficiently different from nonmagnetic white dwarfs to explain the
lack of pre-MCVs as an observational selection effect.  An alternative
explanation is therefore required.

If the origin of magnetic fields in white dwarfs were independent of
their binary nature we would expect the distribution of field
strengths amongst isolated white dwarfs, white dwarfs in
non-interacting binary stars and white dwarfs in cataclysmic variables
to be similar.  This is not the case.  There are no HFMWDs in
non-interacting binary stars in which the companion is a K or~M dwarf.
Such systems, if close enough, would be the progenitors of MCVs.  Ten
per cent of isolated, or single star, white dwarfs are highly magnetic
and two and a half times this fraction of white dwarfs in interacting
cataclysmic variables are highly magnetic.  The fact that there
appears to be no HFMWD in a binary system that has not interacted
suggests that they were once all binary\footnote{\citet{mcclure1997}
made a similar assertion about the R~stars after an extensive radial
velocity search revealed that none has a companion.  If they formed by
an isolated single star mechanism they ought to be equally as common
in wide binary stars as among single stars.}.  The fact that
HFMWDs are common in cataclysmic variables but absent in similar but
wider non-interacting systems suggests that the generation of the
field is entwined with the formation of the cataclysmic variables
themselves.

Here we revisit the proposal of \citet{tout1995} that the strong
magnetic fields are actually generated in the common envelope
evolution that is responsible for shrinking the binary orbit from
giant to dwarf dimensions \citep{paczynski1976}.  In this scenario it
is the systems that merge or just avoid merging that are expected to
have the strongest fields.  These emerge from CE evolution either as
single highly magnetic cores of rapidly spinning giants, which go on
to lose their envelopes and cool as the isolated HFMWDs, or almost in
contact, perhaps already transferring mass, as magnetic cataclysmic
variables.

\section{The observations}

The SDSS sample of $1{,}253$ spectroscopically-observed WD+M pairs
includes only stars with signal-to-noise ratio of at least $5\,$per
pixel.  Any magnetic field over $2-3\,$MG should be detectable by
Zeeman splitting of the Balmer lines.  Two candidate magnetic white
dwarfs were identified with slightly broad lines indicative of perhaps
a $1-2\,$MG field.  However, follow up spectrophotometry with the
Hobby-Eberley Telescope has been obtained for one of these
(WD~0828+4217) and did not confirm that it is magnetic (K. Williams,
private communication).

The SDSS survey also resulted in a large increase in the number of
known isolated magnetic white dwarfs which has now grown to over 170
\citep{vanlandingham2005}. In the most recent compilation by
\citet{kawka2007} 149~magnetic white dwarfs are listed with $B =
3\,$MG or larger.  This is a conservative lower limit of the number of
stars in which the Zeeman splitting is clear enough for a good
spectrum to permit the identification of a HFMWD.

\subsection {Assessing the Statistics of WD+M pairs} 

We can estimate how many of the detached magnetic white dwarfs should
be expected to have a companion by making the assumption that WD+M
pairings occur with the same frequency as for nonmagnetic white
dwarfs.  The sample of white dwarfs within $20\,$pc
\citep{holberg2002} has probably been searched for companions more
thoroughly than any sample of more distant objects.  Of the 109~white
dwarfs, 21~objects or $19\pm 4.5\,$per cent have main-sequence
companions.  This leads us to expect that $14-24\,$per cent of the
149~strongly magnetic white dwarfs, some~$21-35$, should have
nondegenerate companions.  In an unbiased selection of white dwarfs,
$N$ of which should be highly magnetic, we expect to find
$N\pm\sqrt{N}$ which are.  The absence of any main-sequence stars
paired with a HFMWD in this sample is therefore at a four sigma level of
significance.

We can also use larger, more distant samples to improve on these
statistics.  Many of the several hundred hot white dwarfs found in the
Palomar Green Survey \citep{green1986} show the existence of a
companion in the optical spectrum.  \citet{holberg2005} used the Two
Micron All Sky Survey detections at $1.2-2.0\,\mu$m JHK bands to look
for excesses over the white dwarf Rayleigh-Jeans tail.  They found
that $23\,$per cent of the PG sample had definite and $29\,$per cent
had definite or probable cool companions.  If we assume that the
149~strongly magnetic WDs should be a sample with a binary frequency
similar to the Palomar Green Survey, 34 to~43 should have companions.
We now encroach on a six sigma level of significance.

\subsection{Selection effects owing to differences in mass}
\label{masses}

Decided differences are apparent in the fraction of white dwarfs that
are magnetic between the apparent magnitude-limited and the volume-limited
sample.  In the Palomar Green sample, an example of the former, only $2\,$per cent are
strongly magnetic.  In the \citet{holberg2002} sample 21~of the~109
local white dwarfs are magnetic, $19\pm 4\,$per cent.  The latter is arguably
close to volume-limited because all lie within $20\,$pc according to best
distance estimates.  This distinction has been discussed by
\citet{liebert2005} and by \citet{kawka2007}.  Both groups argue that the
likely true frequency of strong magnetism in HFMWDs approaches or
exceeds $10\,$per cent. However, given that the SDSS is magnitude limited, we
can conservatively estimate that we should expect about
$25$~pre-magnetic cataclysmic variables in the sample of the $1{,}253$~stars so far
observed but none have been found.

A systematic effect that might have
gone some way towards explaining this discrepancy is the evidence that
magnetic white dwarfs tend to be more massive and hence less luminous
than nonmagnetic white dwarfs.  \citet{liebert1988} first summarised the 
evidence that several
nearby magnetic white dwarfs with trigonometric parallaxes have
relatively small radii and anomalously high masses and lie below
the sequence of most white dwarfs in an HR Diagram.  These objects 
include the well known Grw+$70^\circ$~8247, G~227-35, G~240-72 and GD~229.
Since then many more magnetic white dwarfs have been shown to be
massive.  However there is also evidence that many of them have more
ordinary masses near $0.6\,M_{\odot}$ or less.  The presence of a very
strong field generally prevents any direct measurement of the mass
through $\log g$ so that the mass estimates for magnetic white
dwarfs have been possible for only a small subset of the known objects.

Three methods have been used to estimate masses for suitable magnetic
white dwarfs.  First, if the field strength is of order $15\,$MG or
less, standard broadening theory applied to each Zeeman component
yields an approximate surface gravity \citep[see examples
in][]{bergeron2001}.  Secondly the measurement of
a good-quality trigonometric parallax is a key way to measure the
radius and luminosity of the magnetic star in comparison with the
nonmagnetic white dwarfs.  A third method has been applied to
binary systems with a nonmagnetic DA paired with a magnetic white
dwarf.  The spectrum of the magnetic component must be subtracted out
if the binary is spatially unresolved.  The fitting of the Balmer
lines of the nonmagnetic object to determine $\log g$ sets the
distance to the system and allows comparison of the radii between the
two components.  Thus the mass estimates from these methods are not as
accurate as those obtained for nonmagnetic objects with $\log g$.

\begin{figure*}
\epsfxsize=0.8\textwidth
\epsfbox{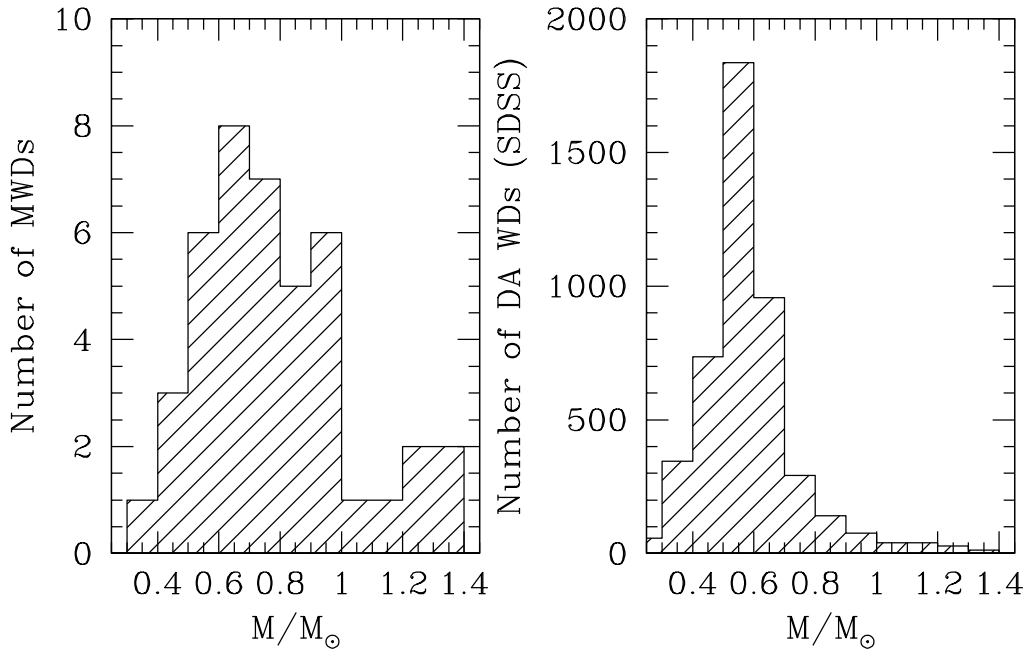}
\caption[]{The distribution of measured masses of magnetic white
dwarfs compared with normal DA~white dwarfs taken from 
\citet{kawka2007} compared with normal DA white dwarfs in the SDSS
sample \citep[without correction for different sampling at different
masses]{kepler2007}.  Note that the two distributions peak at roughly
the same mass but the distribution for magnetic white dwarfs is much
broader than for the non-magnetics.  We account for this by claiming
that the underlying mass distribution is similar but that the magnetic
distribution is augmented at the low end by extra helium white dwarfs
and at the high end by merged double degenerate cores.  According to
our hypothesis, the latter are more abundant and so skew the
distribution in such a way that the mean mass is increased.}
\label{mass}
\end{figure*}

The distribution of measured masses of HFMWDs
\citep{kawka2007} is shown in Fig.~\ref{mass} and compared with normal
DA~white dwarfs in the SDSS sample \citep{kepler2007}.  The mean mass
of the HFMWDs is $0.78\,M_{\odot}$ if we include the rather low mass helium
white dwarfs and $0.82\,M_{\odot}$ if we exclude these stars.  The
mean mass is somewhat higher than the mean mass of $0.58\,M_{\odot}$ of
all white dwarfs and the radii are therefore typically smaller than
those of nonmagnetic white dwarfs.  However the calculations of
\citet{silvestri2007} show that a much larger mass difference is
required to explain the absence of any magnetic pre-CVs in terms of
such a selection effect.  In addition the distribution of masses of
HFMWDs is broad so still includes a substantial fraction of low-mass
stars.

\subsection{Low Accretion Rate Polars (LARPs)}

There exists a small number of observed high field MCVs that have
accretion rates much lower than expected for a semi-detached system
\citep{webbink2005}.  They are thought to be sufficiently close that
the magnetic field of the white dwarf can capture a weak stellar wind
from the companion.  They have periods ranging from $1.3$
to~$4.39\,$hr and magnetic fields from $42$ to~$65\,$MG or so
\citep{schmidt2005,schmidt2007}.  All have low temperatures, $7{,}500
< T_{\rm eff}/{\rm K} < 13{,}000$ and so have not recently emerged
from a common envelope.  We propose that these, and systems like them,
have emerged from the common envelope as very close pairs but must
still wait for gravitational radiation to bring them close enough for
Roche lobe overflow.  They are the magnetic analogues of the normal
pre-cataclysmic variables like V471~Tau with a period of $6.9\,$hr
\citep{warner1995}.  The narrow field range in which the LARPs have
been found is a selection effect related to the use of
cyclotron harmonics in their discovery in the SDSS sample.  We
expect that LARPs with magnetic fields over the entire intermediate
polar to AM~Her range exist and will be found in the future.

\subsection{Doubly Degenerate Systems}

In addition to the MCVs there are seven binary systems in which one
star has a high magnetic field listed by \citet{kawka2007} in their
appendix.  Four of these have been examined in more detail.
EUVE~J0317--855 is thought to have evolved from a triple system in
which two of the stars merged to form the HFMWD \citep{ferrario97}.
G62--46 shows evidence that it has emerged from a CE phase
\citep{bergeron1993} as two very close white dwarfs, one of which is
highly magnetic and of very low mass, $0.25\,M_\odot$.  Similarly
EUVE~1439+75 is a close system that has probably emerged from a CE
phase \citep{vennes1999} as two close and massive white dwarfs, one of
which is highly magnetic.  G141--2 \citep{bergeron1997} has a mass of
$0.26\pm 0.12\,M_\odot$ and so could only have formed in binary
interaction.  Thus there is no evidence that any of these HFMWDs with
degenerate companions must have formed without binary interactions.

\section{Common envelope evolution and magnetic field generation}

Because the white dwarfs in cataclysmic variables must have once been
the cores of giants their binary orbits must have shrunk
substantially from at least several hundred solar radii, to
accommodate a giant, to only a few, so that the red dwarf companions
to the white dwarfs now fill their Roche lobes.  The process leading
to this is not understood at all well but is encapsulated in the
common envelope (CE) evolution described by
\citet{paczynski1976}.  When a giant star fills its Roche lobe,
unstable mass transfer can lead a state in which the giant envelope
surrounds the two dense cores, its own degenerate core and its
companion.  This companion is most likely an unevolved lower-mass
main-sequence star but might itself be already a white dwarf.  These
two cores are then supposed to spiral together inside the CE while
energy and angular momentum are transferred from their orbit to the
envelope which is gradually ejected.

As the cores get closer together their orbital period falls and this
sets up differential rotation within the CE.  By its giant nature the
CE is expected to be largely convective.  Differential rotation and
convection are the key ingredients of a stellar magnetic dynamo
\citep{tout1992}.  \citet{regos1995} go so far as to say that this
dynamo actually drives the transfer of energy and angular momentum
from the orbit to the envelope as well as the strong wind that expels
the envelope.  Irrespective of this, we expect that, at the end of the
common envelope evolution, either when the spiralling cores coalesce
or when all the envelope is driven away, there is a very strong
magnetic field in the vicinity of the hot degenerate core.  This field
can penetrate the nondegenerate surface of the core and become frozen
in as it later cools and contracts.  The closer the cores at the end
of CE evolution the greater the differential rotation in the CE and so
the stronger the expected frozen in magnetic field.

We then expect the strongest white dwarf magnetic fields to form in
the cores of systems that merge during CE evolution.  A main-sequence
companion is likely to disolve into the giant envelope when it has
spiralled in deep enough that its density is comparable with its
surroundings.  The spin angular momentum remaining in the envelope
depends on the details of the CE process as well as the initial
conditions of the system.  If we assume that the remaining envelope
has the specific angular momentum of the original orbit its spin
period would have reduced from years to days.  The degenerate core
therefore finds itself at the centre of a rapidly spinning giant to which
its spin is likely to be coupled.  Because of the small size of the
core its moment of inertia is negligible compared with that of the
remaining envelope.  Such a giant would itself generate a strong
dynamo and spin down quite quickly, typically within $10^4-10^5\,$yr
\citep{tout1992}.  Thus, except in the rare case that the envelope is
almost completely ejected when the cores merge, we would not expect
the HFMWDs to be rapidly spinning by the time they emerge from the
giant envelope.  This is consistent with the tendancy for HFMWDs to be
extremely slow rotators, some with spin periods up to $100\,$yr
\citep{wickramasinghe2000}.  Then, from the common envelope
systems that almost merge, we expect a range of relatively high
magnetic field white dwarfs in MCVs which emerge from the CE very close to
interacting, the polars and intermediate polars, with a corresponding
dearth of such fields amongst the single stars \citep{koester2001}.
Systems which emerge with wider separations should tend to have much
lower fields.  We note at this point that, while we do not understand
the precise mechanism of common envelope evolution, we must expect a
range of magnetic fields associated with any given final separation.
This is then consistent with the fact that the longest period polars
tend to have high fields.  Indeed they must if the field is to be
strong enough to lock the white dwarf spin to the orbit at all.

\section{Consequences}
 
The absence of evidence of any relatively wide pre-magnetic CVs in
some $1{,}253$~WD+M pairs that have so far been studied suggests that
a MCV must be born preferentially in a semidetached state, as an
intermediate polar or AM~Her system, or sufficiently close to
contact with its Roche lobe for the HFMWD to capture the wind of its
companion so that it appears as a low accretion rate polar.
A common envelope magnetic dynamo leads naturally to a scenario in which
this occurs.  In this model systems that emerge more tightly bound
following the CE phase are the systems with the stronger magnetic
fields.

A consequence of this is that the MCVs might be expected to have
intrinsically hotter white dwarf primaries as the systems first come
into contact.  However white dwarfs cool in about $10^7\,$yr to
effective temperatures of about $15{,}000\,$K typical of currently
observed MCVs.  At a period of $2\,$hr the gravitational radiation
timescale is about $3\times 10^9\,$yr.  Indeed it is long enough to
cool to the observed effective temperatures of the LARPs.  The
systematically higher effective temperatures seen in the white dwarfs
in nonmagnetic CVs are attributed to the relative importance of
compressional heating in these systems owing to their higher mean
accretion rates \citep{sion2004}.
 
We note that there are some differences in the properties of
the isolated HFMWDs and the MCVs which have previously been attributed
to their different origins.  First the incidence of magnetism in CVs
(about $25\,$per cent) is significantly higher than in the isolated
white dwarfs (about $10\,$per cent).  This has also been attributed to
possible selection effects \citep{warner1995} though no detailed studies have
been carried out to test this hypothesis.  Secondly the
polars are deficient of white dwarfs with fields in excess of
$10^8\,$G compared with the isolated HFMWDs.  Our hypothesis accounts
for both these differences straightforwardly.  The highest fields are
expected in the CEs in which the cores are closest together.  Thus the
isolated HFMWDs can generally have much higher fields than the MCVs
because the differential rotation in their progenitor CE can be much
greater.  It is greatest just before merging.  The MCVs then fall in a
small range of core separations bordering on the systems that merge.
Wider core separations and the end of the CE phase end up as the
pre-cataclysmic non-magnetic variables and wider systems that will
never interact.  These coupled with single stars and binary stars that
never enter a CE phase make up the low-field white dwarfs in
cataclysmic variables, wider binary stars and single isolated white
dwarfs.

The distribution of the masses of the isolated HFMWDs
(Fig.~\ref{mass}) appears to be made up of a distribution similar to
that of the low-field dwarfs augmented from about $0.6\,M_\odot$
upwards and especially so at the very high masses.  Under our
hypothesis these are simply explained as the result of a common
envelope with two degenerate cores.  Typically this is the second
CE phase in a system with two stars that can both evolve to become
giants.  The first phase leaves a closer MS+WD system.  The second star
then evolves and unstable mass transfer leads to the second CE phase
in which the giant-like envelope surrounds two degenerate cores.  If
the two cores merge to form a massive WD it has a high magnetic field
in accordance with our hypothesis.  If the total mass exceeds the
Chandrasekhar limit the cores may undergo accretion induced collapse
to leave a highly magnetic neutron star.  In either case we expect
accretion during the merging to be fast enough to burn any material
non-degenerately to oxygen and neon \citep{martin2006}.  These stars
ought to emerge rapidly spinning but should also spin down rapidly by
magnetic braking.  One star, EUVE~J~0317-855 \citep{ferrario97}, shows
both a high spin, $P = 12\,$min, and a high mass, $M=1.35\,M_{\odot}$.

The magnetic dynamo model for the origin of fields in MCVs should be
contrasted with the fossil field model that has been used to explain
the properties of the isolated HFMWDs \citep{wickramasinghe2005}.
According to this hypothesis the magnetic flux in the core of the star
that becomes the white dwarf is closely related to the flux of its
main-sequence progenitor.  The mechanism by which this correspondence
occurs has been unclear but recent calculations \citep{zahn2007}
confirm that a fossil field in radiative regions could act as a seed
field to generate a strong magnetic field by a dynamo in the
convective core that becomes the white dwarf \citep[c.f.][]{tout2004}.
An alternative dynamo origin in single stars was proposed by
\citet{levy1974} but it still relies on the intrinsic stellar spin of
a single star.  The major difference in our hypothesis is that, unlike
in single star evolution, strong differential rotation is an {\it
essential} characteristic of common envelope evolution.  Our
assumption is that any intrinsic fossil field that is present is
destroyed or at most serves as a seed field for the magnetic dynamo in
the CE phase.

The space densities and observable lifetimes of CVs and white dwarfs
are not well known but we can use current estimates to check
consistency with our hypothesis.  There are $1.1\pm 2.3\times 10^{-5}$
CVs per cubic parsec \citep{pretorius2007} and about one quarter of
these have high fields.  There are $3\times 10^{-3}$ white dwarfs per
cubic parsec \citep{liebert2005} and about one tenth of these are
HFMWDs.  If we assume that the observable lifetime of a CV, the time
over which its mass-transfer rate is sufficiently high, is about one
tenth that of an isolated white dwarf, the time for it to cool below
detectable limits, then the birth rate of HFMWDs is about three times
that of CVs.  Thus three times as many systems entering a CE phase
should end up merging as emerge separated but close enough to become a
CV.
  
\section{Conclusions}
  
The fact that no white dwarf with a surface magnetic field over
$3\,$MG has been found in a detached binary system suggests that all
such highly magnetic white dwarfs have a binary origin.  If half of
the stars in our neighbourhood have a binary companion and half of
these are sufficiently separated not to have interacted then there
should be at least one quarter as many magnetic white dwarfs in wide
detached binary systems as appear as single stars unless their origin
depends on binary interaction.  This is not the case and so argues
very strongly against any single star evolutionary origin of HFMWDs
and very much in favour of a mechanism that relies on binary
interaction.

Given the conundrum, relating to the MCVs,
that there is an absence of evidence for any pre-MCVs in some $1{,}253$
WD+M pairs that have so far been studied, we have provided the
solution that the origin of high
magnetic fields in white dwarfs relies on a magnetic dynamo during the
common envelope phase of binary evolution.  Systems with the strongest
magnetic fields emerge from the CE phase either as merged single stars
or with their secondary stars
more nearly in contact with their Roche lobes thus reducing their
chance of being detected as pre-magnetic cataclysmic variables.
 
Our hypothesis also predicts the existence of high field magnetic white
dwarfs that result from systems that merge during the common envelope
phase.  The absence of any MCVs in detached binary stars leads us to
conclude that all highly magnetic white dwarfs have formed in this
way.

\section*{Acknowledgements}
We thank Gary Schmidt and Adam Burrows for enlightening discussions.
CAT thanks Churchill College for a Fellowship.  DTW is grateful to the
Institute of Astronomy, Cambridge for hospitality.  JL acknowledges
support from the U.S. National Science Foundation, grant AST 03-07321,
for work on white dwarfs in the Sloan Digital Sky Survey.

\label{lastpage}
\end{document}